# Conductor Properties and Coil Technology for a Bi2212 Dipole Insert for 20 Tesla Hybrid Accelerator Magnets

E. Barzi*

*Fermi National Accelerator Laboratory, Batavia, IL 60510, USA, and Ohio State University, Columbus, OH 43210, USA*

ABSTRACT

Developing HTS dipole inserts producing fields larger than 5 T within 15 T $Nb_3Sn$ outserts is necessary to generate 20 T or higher fields for future high energy colliders. Dipole inserts based on the cos-theta coil geometry with various stress management concepts and Bi2212 superconducting strand and cable are being developed at Fermilab both within and beyond the U.S.-MDP effort. The ultimate goal is to develop coil technology and an approach to manage azimuthal and radial strains of high temperature superconductor inserts when integrated within $Nb_3Sn$ outserts as a hybrid magnet system. This white paper reviews Bi2212 conductor properties and coil technologies, and proposes new ideas to face the challenges that Bi2212 still presents as an accelerator magnet conductor.

## 1. EXECUTIVE SUMMARY AND RECOMMENDATIONS

### 1.1. Summary

The HEP global community has been ushering in a new era of high-tech accelerator development through strong endorsement of the 2021 European Strategy for Particle Physics of "high-field superconducting magnets, including high-temperature superconductors."* CERN's Magnet Division has been investing $100M+ over 5 years on this topic for M&S alone. Within the U.S. Magnet Development Program (US-MDP), a key task at FNAL is that of developing high temperature superconducting (HTS) inserts producing fields larger than 5 T within 15 T outserts made of $Nb_3Sn$ to generate 20 T+ for future circular colliders. Whereas the priority for $Nb_3Sn$ accelerator magnets is that of significantly reducing or eliminating training, Bi2212 is a most promising conductor to achieve higher magnetic fields. However, there are still several challenges that still lie ahead for Bi2212 to be used in hybrid magnets. This white paper reviews Bi2212 properties and coil technologies, and proposes new ideas to face the technical challenges in Bi2212 materials and accelerator magnet technology, including the loss of critical current density in magnets with respect to wires and cables (magnet degradation).

### 1.2. Recommendations

- Heat treating Bi2212 at high gas pressure does not eliminate Bi2212 leaks from the conductor. Leaks are unacceptable in accelerator magnets due to the risk of shorts, for instance. Also, systematic leaks of melted Bi2212 phase are a major reason for the loss of critical current in coils. It is necessary to work in collaboration with OST-Bruker and any other capable industry to design billets that are adequate for Rutherford cabling. Design modifications, such as increasing the outer Ag barrier thickness, decreasing fill factor, and reoptimizing the subelement geometry, will sacrifice real estate in the conductor, and hence reduce the critical current of the round wire. However, this decrease will be compensated with much smaller degradation of the critical current in the magnet itself.

---

*https://indico.cern.ch/event/905399/contributions/4099237/attachments/2259137/3834090/LHCP-Gianotti.pdf

- To realize an effective accelerator magnet an insulation material chemically compatible with Bi2212 and its high temperature processing in oxygen has still to be found. Existing methods using mullite sleeve and $TiO_2$-polymer slurry have not shown to completely eliminate leaks. Further investment in research and development of compatible insulation is sorely needed.
- In order to use Bi2212 coils as inserts in very high field magnets, it is critical to control and limit their stresses and strains as Bi2212 is more sensitive to stress than $Nb_3Sn$. This can be done by acting on the following fronts:
  1. Invest in research and development of methods to mechanically reinforce the wire Ag matrix and/or the Rutherford cable itself;
  2. Use coil stress management elements to reduce stress in the insert coils while also being chemically compatible with the Bi2212 processing;
  3. Verify progress by accurate critical current measurements of Bi2212 cable samples under transverse pressure.
- To lower costs and simplify the processing of Bi2212 inserts for hybrid accelerator magnets, it is worth reconsidering a Split Melt Process (SMP) in which the overpressure heat treatment of the Bi2212 is split into two separate heat treatments and the coil is wound between them. This would simplify temperature control for large magnets at the most sensitive temperature step, promote more homogenous Oxygen diffusion, and possibly prevent or reduce leaks. A hybrid approach where one or both cycles of the Split Process are carried out at 50 bar was never explored experimentally and should be.

## 2. INTRODUCTION

The 2014 Particle Physics Project Prioritization Panel (P5) strategic plan for U.S. High Energy Physics (HEP), echoed by the 2015 HEPAP subpanel review of the General Accelerator R&D (GARD) program, endorsed a continued world leadership role in superconducting (SC) magnet technology for future Energy Frontier facilities. Developing accelerator magnets up to or exceeding 20 T requires both High Temperature (HTS) and Low Temperature Superconductors (LTS).

The continued progress toward higher magnetic fields holds significant potential for general advances in science and technology [1]. For the past twenty years there have been steady increases in the field strength of SC magnets thanks to constant advances in $Nb_3Sn$ conductor and magnet technology [2]. In 2009, Internal tin $Nb_3Sn$ conductors have enabled a commercial 23.5 T magnet for 1 GHz NMR spectroscopy [3], with an actively shielded version produced in 2016 [4]. On May 4[th], 2020, Bruker Corporation announced the successful installation and customer acceptance of the world's first Avance™ NEO 1.2 GHz NMR system at the University of Florence, Italy [5]. Bruker's NMR magnets with 54 mm bore utilize a novel hybrid technology with HTS in the inner sections, and LTS in the outer sections. In the accelerator arena, following the previous 1997 record magnetic field at 4.5 K of 12.8 T (13.5 T at 1.9 K) in the four-layer cos-theta D20 LBNL dipole, in 2019 the FNAL team achieved a world record field at 4.5 K of 14.1 T in an accelerator dipole [6], pushing it up to 14.6 T at 1.9 K in 2020 [7].

The work on Bi2212 conductor and technology started in the U.S. several years ago [8] and now continues through the U.S. Magnet Development Program (US-MDP) [9]. US-MDP was established in 2016 to answer the 2014 P5 call for its strategic plan for HEP. A major goal of the US-MDP SC magnet program is that of developing HTS inserts producing fields larger than 5 T within 15 T Nb$_3$Sn outserts to generate 20 T or higher fields for future high energy colliders [10].

Practical and commercially available HTS include Bi2212 round wires and REBCO tapes. Progress with round Bi$_2$Sr$_2$CaCu$_2$O$_{8-x}$ (Bi2212) composite wires, which can be used to produce Rutherford cables, makes them particularly suited for use in high-field accelerator magnets [11], [12]. However, Bi2212 coil performance is still inferior to the short sample limit expected from the wire performance. Therefore, a number of challenges still has to be addressed. Composite Bi2212 is a soft and very delicate material which needs rigorous empirical laws for Rutherford cabling to minimize internal damage. Once the cable is formed and used to wind a coil, the Bi2212 coil requires a multistage heat treatment in a highly corrosive Oxygen atmosphere at maximum temperatures close to 900ºC. Temperature homogeneity has to meet stringent gradient specifications. In accelerator magnets wound inside a mechanical structure, oxygen access to the superconducting volume might be limited and requires appropriate solutions. Also, the Bi2212 leaks that were observed at 1 bar reaction in Oxygen (O$_2$) do not desist when performing the 50 bar heat treatment. Spots or discolorations form where Bi2212 liquid leaks through the encasing Ag alloy metal at high temperatures and reacts with surrounding materials, including the insulation. Most leakages in small racetracks occur at the Rutherford cable edges, and they degrade the $J_e$ locally [13]. This problem is related to both the strand ability to withstand deformation and the chemical compatibility of the insulation material with the Ag alloy during heat treatment in O$_2$.

Similarly to Nb$_3$Sn, Bi2212 wires and cables are sensitive to strain [14] - [17]. Although Bi-2212 is universally made with the Powder-in-Tube (PIT) technique, it appears that strain behavior depends on the Ag alloy used by each manufacturer. Whereas there exist very accurate measurements on tensile strain, it is rather the azimuthal pressure, i.e. transverse to the cable, that is one of the main stress components in accelerator magnets. More data on Bi2212 performance sensitivity to cable loading are vital. It is already certain, however, that stress management concepts will need to be applied to insert coils' designs when aiming at large magnetic fields of hybrid magnet systems. Coil test, quench detection and protection also need to be addressed.

A Bi2212 dipole insert for an accelerator magnet is made of insulated Bi2212 Rutherford cables wound inside a mechanical support structure. This applies for instance to both the Canted Cosine Theta (CCT) coil concept [13] by the LBNL group in collaboration with ASC-NHMFL-FSU, and to Fermilab's coil inserts based on the traditional cos-theta coil design with stress management elements [18]. To design an insert in this latter geometry, the maximum aperture is given by the mechanical properties of the Bi2212 Rutherford cable, which has to be bent in the plane of its flat face at the coil mid-plane, and in a plane perpendicular to it at the poles. Refined magnetic and mechanical analyses are performed with ROXIE and ANSYS. Parts for the insert coil can be designed using BEND and can be fabricated by 3D sintering technology. As a first step, practice coils using "dummy" Nb$_3$Sn cables and 3D printed plastic parts are wound, impregnated with epoxy and cut to examine turns position in various parts of the coil. After winding, the actual coil can be heat treated (reacted to form Bi2212 stoichiometry) at 50 bar in a furnace at NHMFL, and shipped back for impregnation with epoxy, instrumentation and cold testing at FNAL. In preparation for this and future inserts, cable development and characterization, including transport properties of extracted strands, cable at field, and cable under pressure are performed. This white paper reviews some of these steps and proposes new ideas to face the challenges that Bi2212 still presents as a magnet conductor.

## 3. WIRE, CABLE AND COIL DESIGN

### 3.1. Bi2212 Wire

$Bi_2Sr_2CaCu_2O_{8-x}$ (Bi2212) is one of several copper-oxide high temperature superconductors, which in addition to much higher critical temperatures also have very high critical fields (> 50 T at 4.5 K) compared with LTS. Furthermore, Bi2212 is the only copper-oxide material which can be easily melt processed, which enables it to be fabricated in a wide variety of shapes, including isotropic round multifilamentary wire. To achieve this so-called partial melt processing, a multistage heat treatment in highly corrosive Oxygen atmosphere at very uniform temperatures up to 900ºC is required. Bi-2212 is universally made with the Powder-in-Tube (PIT) technique. Whereas the matrix embedding the filaments is pure Ag, the wire fabrication process typically uses a Ag0.2%Mg alloy outer sheath which is dispersion strengthened by oxidation of the Mg during heat treatment.

Significant advances were made in the development and industrialization of Bi2212. Km-length quantities of Bi2212 composite round wires are commercially produced. The critical current density $J_c$(20 T, 4.2 K) and the engineering current density $J_e$(20 T, 4.2 K) of the round wire increased from 1500 A/mm² and 400 A/mm² or less respectively in the 2000s, to the $J_e$ values shown in Fig. 1[13]. Two are the main factors that led to this improvement:

1. Removal of the 30% porosity in as-drawn Bi2212 wires by an overpressure processing heat treatment (OPHT).
2. The introduction of a new chemical powder technology by nGimat (now Engi-Mat), which produces highly homogenous Bi2212 precursor powders with very good composition control.

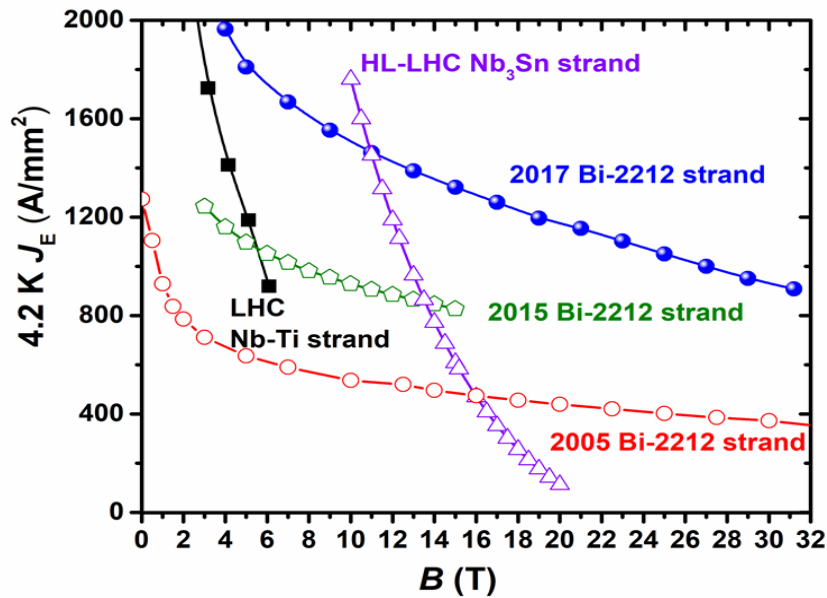

Figure 1: Improvement of the best of their time Bi2212 wire $J_e$ (B) as compared with the LHC NbTi and HL-LHC Nb₃Sn wire specifications [13]. The 2005 curve represents best performance in the 2000s for Bruker-OST wires using Nexans powder; the 2015 curve shows similar wires' improvement in performance when undergoing the OPHT process; the 2017 curve is the best performance obtained by Bruker-OST with the new Engi-Mat powders.

Fig. 2 gives an example of a Bi2212 typical heat treatment cycle, which is used for the partial melt processing at 1 bar of gas pressure. The conductor is heated up to a maximum temperature $T_m$ above the peritectic decomposition temperature of the Bi2212 phase and slowly cooled down to form a well-connected and aligned Bi2212 grain structure.

The gas is either pure $O_2$ or a mixture of $O_2$ and Ar ($O_2$/Ar) with $O_2$ partial pressure of 1 bar. The OPHT process uses 98% Ar and 2% $O_2$ with a total gas pressure at 50 bar, with the $O_2$ partial pressure still at 1 bar.

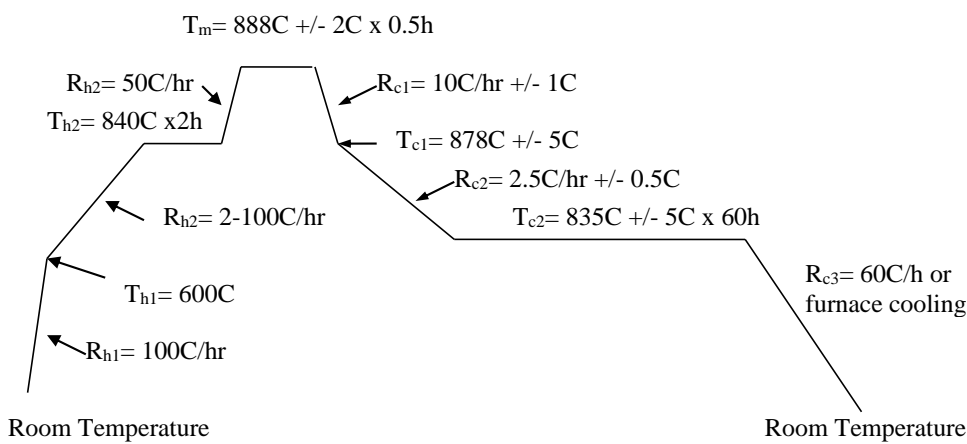

Figure 2: Bruker-OST optimized heat treatment for Bi2212 wire at 1 bar gas pressure.

In 2008 studies were published on splitting the standard Bi2212 heat treatment in two portions for the magnet to be wound between the two cycles [19], [20]. This approach was called Split Melt Process (SMP) to allow React-Wind-Sinter coil fabrication. The very first studies produced promising results on round wires. For a split temperature $T_s=T_{c1}$ in Fig. 2, straight wire samples saw a 40% increase in $J_c$, and a 30% increase was measured in samples bent down to 15 mm bending radius after the first stage of the heat treatment. However, such studies were interrupted when the OPHT process was introduced, since the latter increases $J_c$ by more than 200% or more [21]. Therefore, several parameters associated to the Split Melt Process were not optimized to possibly produce further $J_c$ progress.

A hybrid approach where one or both cycles of the Split Process are carried out at 50 bar was never explored. A Split Melt Process in which the Bi2212 OPHT is split into two separate heat treatments and the coil is wound between them, would offer several advantages. In the assumption that the ceramic leakage and $J_c$ sensitivity to peak times and temperature gradients are more critical in the liquid state [19], either loosely wound wire or Rutherford-type cable would be heat treated at 50 bar for the first stage of the thermal cycle. The fact that high $J_c$ is obtained in Bi2212 within small ranges of peak temperature and time will limit the coil size that can be uniformly heat treated. In the first stage of the Split Melt Process, a spool of Rutherford cable would have a more compact volume than a large coil, with a more precise control of the heat treatment parameters. Another important benefit of heat treating a spool of loosely wound cable as opposed to a tightly wound coil might be that of reducing ceramic leaks that occur at the liquid stage, and of a more homogenous $O_2$ diffusion. Before going through the second stage of the thermal cycle, the coil would then be wound. If the second step could be performed at atmospheric pressure without loss of $J_c$, this would substantially lower the cost of smaller high-pressure furnace equipment. However, even if the second step will have to be performed at 50 bar and produce at least the same or higher $J_c$ than in standard OPHT cycles, this hybrid approach would simplify temperature control for large magnets at the most sensitive temperature ($T_m$) step, promote more homogenous $O_2$ diffusion, and possibly prevent or reduce leaks.

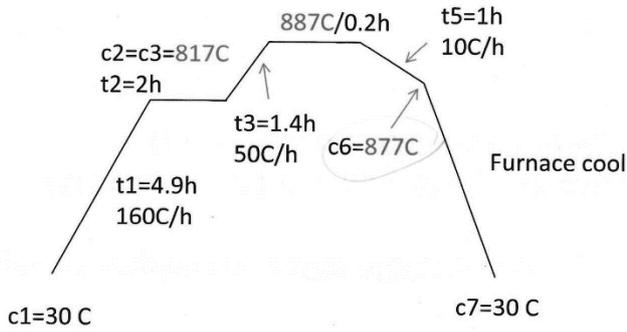 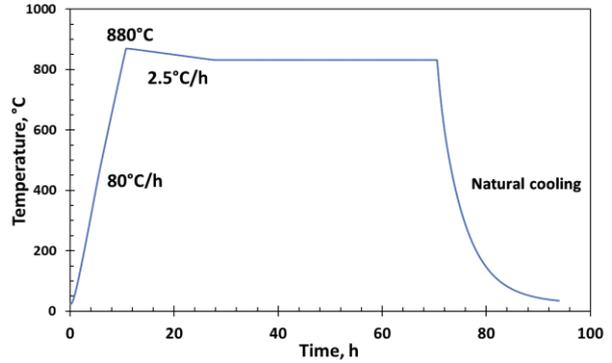

Figure 3: Left: Nominal cycle for the first part of a Split Melt Process (SMP) heat treatment performed at FSU at 50 bar gas pressure. Right: Actual cycle for the second part of the SMP heat treatment performed at FNAL at 1 bar gas pressure.

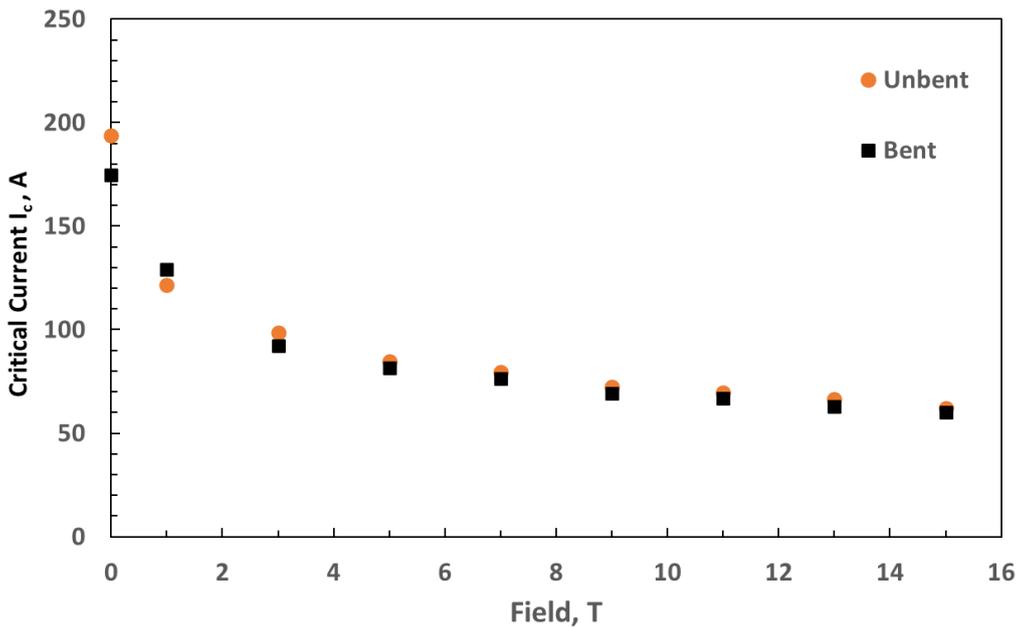

Figure 4: Critical current vs. magnetic field for the Bi2212 undeformed sample and for that which had be wound with a 6.6 mm radius.

Some preliminary SMP experiments were conducted at FNAL using a 0.8 mm Bi2212 wire with low $J_c$. The wire was sent to FSU to execute the first part of the SMP heat treatment at 50 bar of gas pressure (Fig. 3, left). Before heat treating straight samples 6 inch long, the samples were sealed at FSU with Ag following the method described in [22]. Once back at FNAL, some of the pre-treated samples were wound around a metal cylinder of 13.13 mm in diameter before unwinding them and bringing them back to their original straight form. For the second part of the SMP cycle (Fig. 3, right), both undeformed straight samples and previously wound straight samples were used. Unfortunately, this wire leaked in the second part of the heat treatment, and the experiment will have to be repeated with a better quality billet. However, sections of samples without leaks 4 cm long could be used for testing their critical current $I_c$ as function of field. The criterion used for determining $I_c$ was 1 µm/cm. The $I_c$ results are shown in Fig. 4 for the undeformed sample and for that which had be wound with a 6.6 mm radius. If these results will be reproduced also for other Bi2212 billets, they show that indeed coils could be wound after partial heat treatment. The experiment will be performed also with the second part of the SMP cycle at 50 bar gas pressure.

### 3.2. Bi2212 Rutherford Cable

Whereas the best wire performance is usually obtained by $O_2$ pre-annealing of the Ag0.2%Mg sheathed strand, the hardened sheath which results from this pre-anneal severely restricts the diameter around which the wire can be bent without cracking. To enable cabling of Bi2212 round stands, in the past specific $O_2$ an anneal process called PAIR, now discontinued, was used. Nowadays, at Bruker-OST the thermo-mechanical process was re-engineered to enhance wire sheath integrity and avoid leakage during the melt-recrystallizing heat treatment.

However, the present Bi2212 billets fabricated for US-MDP are not suitable for Rutherford cables, which may explain why Bi2212 coils keep leaking. There are design modifications, such as increasing the outer Ag barrier thickness, decreasing the fill factor, and reoptimizing the subelement geometry, that would have to be implemented on the wire to make it adequate for cabling. The US-MDP Conductor Procurement and Research Development CPRD has so far focused on $J_c$, which clearly would have to be a bit sacrificed in stronger wires with less real estate for conductor. Unfortunately, heat treating under high pressure does not eliminate leaks, which are unacceptable in accelerator magnets.

For Rutherford-type cables fabrication, a Bi2212 wire has to pass a nominal bending test, i.e. the wire should be able to bend on its own diameter without cracking. Cable development is performed by designing and fabricating samples of different geometries using state-of-the-art wires, with the purpose of studying the effect of cable parameters and processing on their performance. This includes for instance the sensitivity of electrical properties (critical current $I_c$, residual resistivity ratio RRR) and internal structure (architecture, filament shape and spacing, sheath composition) to cable compaction; measurements of cable stability and AC losses; measurements of 3D cable expansion during the fabrication process and during reaction; effects of intermediate annealing when using a 2-pass fabrication process; etc. Rutherford cable finite element models that evaluate for each considered cable geometry what is the plastic strain seen by the strands during fabrication, what are the most critical strand locations, and that predict local damage whenever the failure mechanisms of a specific strand technology are known, can be used to aid in cable design [23].

### 3.3. Bi2212 Insert Coil Designs

The design of high-field dipole inserts made of Bi2212 is extremely challenging due to the small volume typically available inside the bore of a $Nb_3Sn$ outsert. For a 2-layer insert coil design within a 60 mm diameter space, the inner winding requires a very low radius of curvature and the outer one has the most critical bending issues due to the hard bending component at the mid-plane. The inner region of the inner layer is the most likely to get damaged. The small radius of the wound Bi2212 cables can compromise the magnet during the winding, reaction and/or impregnation processes.

An example of 2-layer insert coil design that uses Rutherford cable is shown in Fig. 6 [24]. The design was based on both length (15 m) and cross-section of an insulated rectangular cable that was provided by LBNL. The actual wire and the cross-section of a similar cable are shown in Fig. 5 and their main parameters in Table 1. The Bi2212 wire was produced by Bruker OST LLC using precursor powder by Engi-Mat. The cable was made and insulated at LBNL. The cable insulation consists of 0.15 mm thick mullite braided sleeve chemically compatible with the superconductor heat treatment in $O_2$.

Because of the large Lorentz forces and the small available volume, the coil support structure has to be designed to protect the strain sensitive Bi2212 cable from mechanical over-compression during assembly and operation. This is done by winding the insulated cable within grooves in the structure. In addition to stress management, this type of support structure provides turn positioning during assembly and operation. In this specific example, the two coil layers are wound on one single structure, the inner layer from inside and the outer layer from outside, as also shown with the plastic model in Fig. 7.

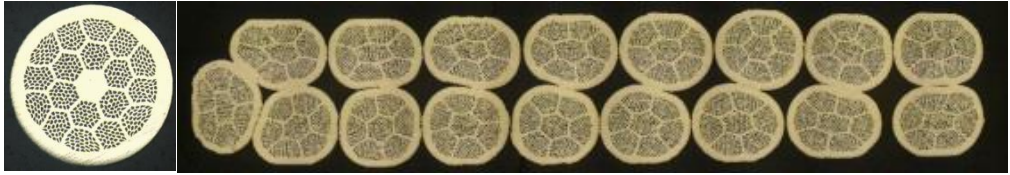

Figure 5: Bi2212 round composite wire (left) and Rutherford cable (right) of same cross-section as that received at FNAL.

Table 1. Bi2212 cable and strand parameters

| Parameter | Value |
| --- | --- |
| Cable ID | LBNL-1110 |
| Number of strands | 17 |
| Bare cable width (mm) | 7.8 |
| Bare cable thickness (mm) | 1.44 |
| Cable transposition pitch (mm) | 58 |
| Billet ID | PMM180207-2 |
| Strand diameter before/after reaction (mm) | 0.8/0.778 |
| Strand architecture | 55 x 18 |
| Strand fill factor (%) | 23 |
| Strand twist pitch (mm) | 25 |
| Strand $I_c(4.2K,15T)$ after NHMFL 1 bar Standard HT (A) | 157-175 |
| Strand $I_c(4.2K,5T)$ after NHMFL 50 bar OPHT (A) | 460-640[*] |

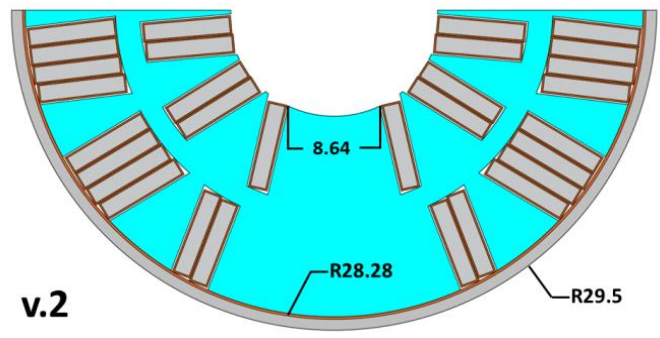

Figure 6: Half-coil cross-sections inside optimized v.2 support structure.

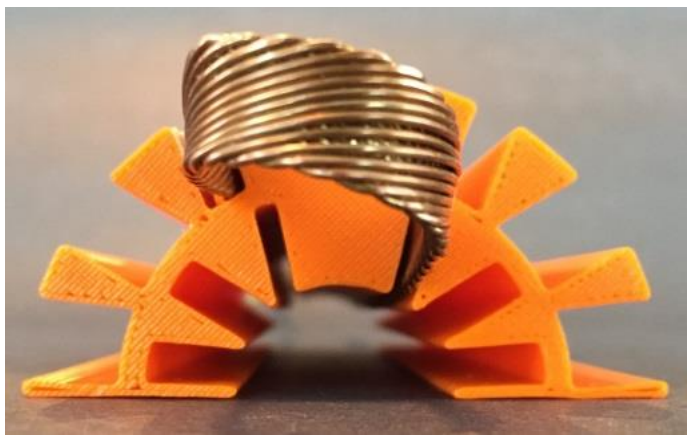

Fig. 7. Plastic model of the coil structure with inner-layer pole turn and interlayer transition turn.

This type of mechanical structure significantly reduces the conductor surface in contact with the oxygen. However, experience by other groups has shown that this does not impact the results of the reaction for coils of up to 20 layers wound with wire [25]. However, this will have to be verified for larger coils made of insulated Rutherford cable.

Whereas Bi2212 is known to expand during reaction at 1 bar, it contracts at 50 bar. Accurate measurements of cable thickness, width and length expansion/contraction will be needed to appropriately design the groove.

## 4. COIL TECHNOLOGY AND MATERIALS CHEMICAL COMPATIBILITY

For the mandrel and parts of the coil structure, at LBNL Aluminum-bronze-954 is used, as opposed to Inconel-600, which is more expensive and challenging to machine. However, Inconel can be used in 3D sintering technology, whereas Aluminum-bronze cannot. When using Aluminum-bronze, every former part is pre-oxidized before the Bi2212 heat treatment, in order to create an $Al_3O_3$ oxidation layer that prevents the material from absorbing $O_2$ from the environment during the coil heat treatment.

Presently, the Bi-2212 Rutherford cable is insulated with mullite braided sleeve, i.e. ceramic braid $2Al_2O_3$:$SiO_2$, and impregnated with epoxy. At LBNL, Mix-61 from NHMFL is sometimes used for this purpose. Other available impregnation materials are CTD-101 and Matrimid. Within an U.S.-Japan Science and Technology Cooperation Program in High Energy Physics titled "High heat capacity and radiation-resistant organic resins for impregnation of high field superconducting magnets", organic olefin-based thermosetting dicyclopentadiene (DCP) resin, commercially available in Japan as TELENE® by RIMTEC Corporation, is also being studied as impregnation material.

Unfortunately, the Bi2212 leaks that were observed at 1 bar reaction in $O_2$ do not desist when performing the 50 bar heat treatment. Spots or discolorations form where Bi2212 liquid leaks through the encasing Ag alloy metal at high temperatures and reacts with surrounding materials, including the insulation [13]. Most leakages in LBNL small racetracks occur at the Rutherford cable edges, and they degrade the $J_e$ locally. An example of leaks in Rutherford cables and their composition can be found in [26]. After reaction, the surface of all cables under study showed black spots embedded in the Ag coating as in Fig. 8, left and center. When tested at 4.2 K and self-fields of 0.1 to 0.3 T, an $I_c$ degradation of about 50% was found for all these cables. This current reduction on the cables was significantly and systematically larger than that of their extracted strands.

SEM/EDS analysis performed on the surface of a cable showed that the composition of this black material was very close to that of Bi-2212 (Spectrum 1 in Table of Fig. 8, right). In a small crater at the edge of the sample, analysis showed several oxide phases, with specific shapes and morphologies, such as Bi-2212 (needle like grains), Bi-2201 (step like grains), (1,0) phase (spherical grains) and others. However, because no leaks were observed on the extracted strands, which performed well, the hypothesis can be made that this problem is not as much related to the strand ability to withstand deformation as to the heat treatment of the cables itself.

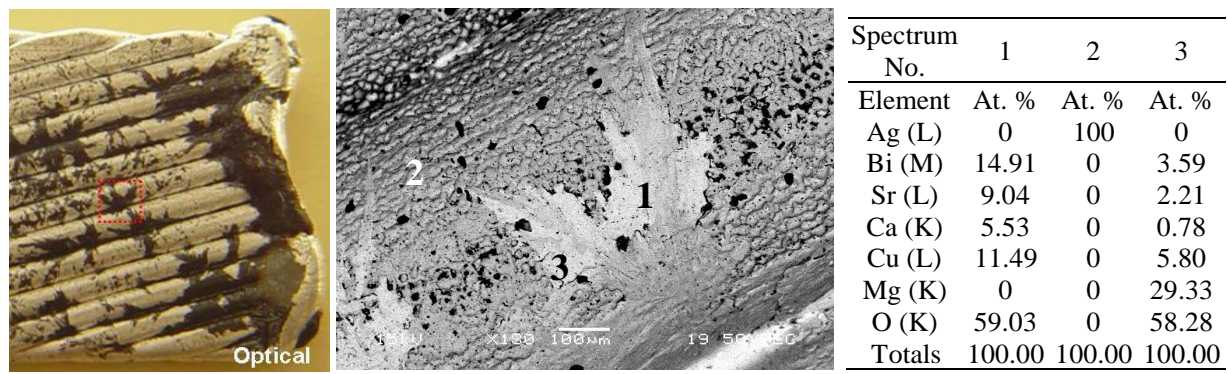

| Spectrum No. | 1 | 2 | 3 |
|---|---|---|---|
| Element | At. % | At. % | At. % |
| Ag (L) | 0 | 100 | 0 |
| Bi (M) | 14.91 | 0 | 3.59 |
| Sr (L) | 9.04 | 0 | 2.21 |
| Ca (K) | 5.53 | 0 | 0.78 |
| Cu (L) | 11.49 | 0 | 5.80 |
| Mg (K) | 0 | 0 | 29.33 |
| O (K) | 59.03 | 0 | 58.28 |
| Totals | 100.00 | 100.00 | 100.00 |

Figure 8: Bi-2212 Rutherford cable after reaction (left), back scatter image of circled black spot (center), and composition Table in marked locations (right) [26].

As an attempt at reducing Bi2212 leaks, before insulating the Rutherford cable with the mullite sleeve, at LBNL the cable is brushed with a thin layer of $TiO_2$-polymer slurry, i.e. $TiO_2$ powder mixed with ethanol using various ratios [25]. During coil winding this coating is also applied to the insulated cable onto the mullite sleeve. Once the insulated cable is painted it has to be wound promptly because the curing time of the slurry is of only about 10 minutes [25]. This method does not eliminate leaks, but should at least reduce their number. Further investment in research on compatible insulation might solve this problem permanently.

In the meantime, a small effort is being invested at FNAL to reproduce this method while at the same time identifying the minimum bending radius that is acceptable for a Bi2212 cable insulated with mullite sleeve and coated with $TiO_2$ slurry and heat treated in Oxygen within an Inconel structure.

For the design of the reaction tooling, thermal properties are necessary (see Table 2).

Table 2. Thermal contraction and cold elasticity modulus of Bi2212 and $Nb_3Sn$ coil and mirror/dipole structural materials

| | Structural element | Material | Thermal contraction (300K -4K), mm/m | Elasticity modulus (cold), GPa |
|---|---|---|---|---|
| HTS insert coil | Coil | Bi2212 | 2.9/3.3 (rad/azim) | 25/18 (rad/azim) |
| | Structure | Al bronze/Inconel 718 | 3.6/2.4 (rad/azim) | 120/219 (rad/az) |
| LTS outsert coil | Coil | $Nb_3Sn$ | 2.9/3.3 (rad/azim) | 40/40 (rad/azim) |
| | Poles/wedges | Ti-6Al-4V | 1.7 | 125 |
| Mirror/dipole structure | Yoke | Iron 1045 | 2.0 | 225 |
| | Clamp | Aluminum | 4.1 | 81 |
| | Skin | 304L | 2.9 | 210 |

## 5. COIL PERFORMANCE ANALYSIS

Some Bi2212 half-coils could be tested both individually and in the background field of $Nb_3Sn$ coils. The latter options include an 11 T mirror dipole structure HFDM [27], and a mirror configuration for the future 17 T 4-layer $Nb_3Sn$ coil [28]. Finally, a whole dipole insert could be tested inside the entire 17 T dipole. A coil short sample limits can be estimated from the dipole load line, using the minimal value of strand $I_c$ from Table 1 and a typical $I_c(B)$ parameterization for Bi2212 wires as critical surface. Fig. 9, left, shows the $I_c$ limits as function of the outsert magnetic field. Fig. 9, right, shows the corresponding field produced by the insert as function of the outsert magnetic field. In both plots, the $I_c$ cabling degradation is used as a varying parameter. For the insert used as an example, its maximum field in the aperture at $B_{ext}=0$ would be within 3.4-4.5 T and at $B_{ext}=15$ T it would reduce by only ~1 T.

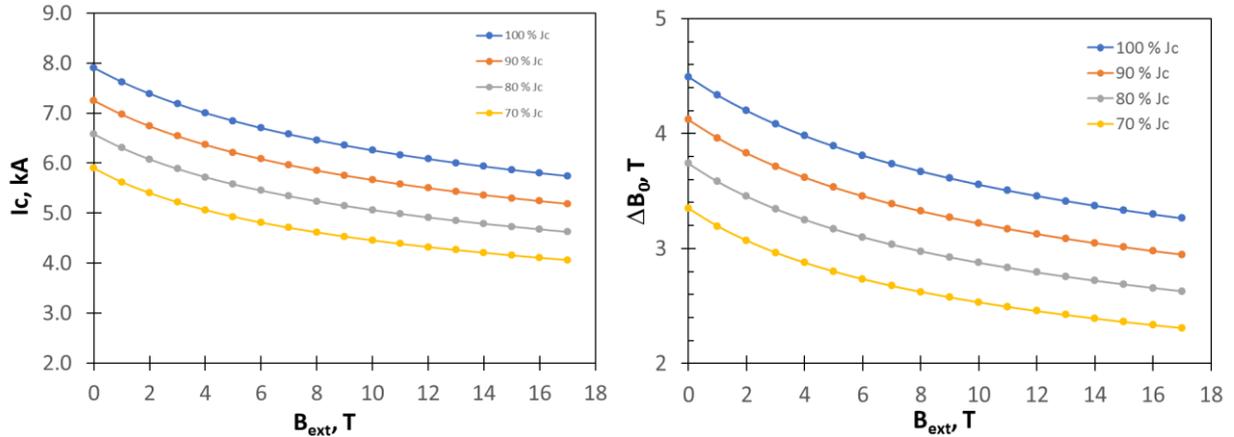

Figure 9: Left: Bi2212 insert $I_c$ limit at various external magnetic fields and conductor degradation levels. Right: Expected maximum field in the coil insert aperture at various external fields and $I_c$ degradation levels in coil.

### 5.1. Coil Force Distribution

For the dipole insert used so far as an example, the Lorentz force values and directions in the coil blocks without an outsert present are shown in Fig. 10 [18]. The forces in the inner layer are practically horizontal and in the outer layer there are both force components. In an external dipole field the horizontal component of the Lorentz force will proportionally increase whereas the vertical one will not change. Calculated values of maximum equivalent stress in the Bi2212 dipole insert used as an example and in an 11 Tesla $Nb_3Sn$ outsert are shown in Table 3 for both the case of Al-bronze and Inconel-718 structural elements. In Table 3 stress values are shown for both cases of testing in a mirror (half) and whole magnet configurations.

Table 3. Max. equivalent stress in Bi2212 and $Nb_3Sn$ coils and coil structural elements (MPa)

| Structural element | Material | Mirror | | | | Dipole | | | |
|---|---|---|---|---|---|---|---|---|---|
| | | Al Bronze | | Inconel 718 | | Al Bronze | | Inconel 718 | |
| | | 0 kA | 8 kA | 0 kA | 8 kA | 0 kA | 8 kA | 0 kA | 8 kA |
| Insert structure | | 283 | 248 | 895 | 829 | 260 | 150 | 588 | 476 |
| Insert coil | Bi2212 | 116 | 92 | 96 | 104 | 135 | 78 | 103 | 70 |
| Outsert coil | $Nb_3Sn$ | 102 | 121 | 98 | 114 | 114 | 124 | 113 | 124 |
| Outsert pole 1 | Ti Alloy | 487 | 157 | 445 | 131 | 620 | 183 | 551 | 142 |
| Outsert pole 2 | Ti Alloy | 266 | 193 | 269 | 197 | 257 | 166 | 262 | 170 |

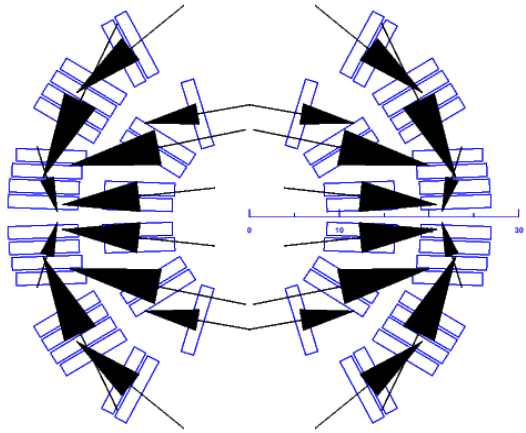

Figure 10: Left: Lorentz forces in coil blocks [18].

## 5.2. $I_c$ Strain Sensitivity

As shown above, the main stress components on the coil conductor are transverse to the Rutherford cable length. The largest stresses are both transverse to the cable face and to the cable edge. Bi2212 wires and cables are sensitive to strain [14] - [17], and whereas there exist very accurate measurements on tensile strain, it is rather the azimuthal pressure, i.e. transverse to the cable, that is one of the main stress components in accelerator magnets. More data on Bi2212 performance sensitivity to cable loading are vital.

### $I_c$ Sensitivity to Tensile/Compressive Strain

Measurements of $I_c$ axial strain sensitivity of superconducting wires are performed with a helical Walters' spring device, for applying axial strain to the sample [29]. The FNAL setup is equipped with two concentric Cu tubes that are used both to transport up to 2 kA of current and also as 60 N·m torque carriers [30]. The spring is typically made of cold-worked and precipitate-hardened Cu-2%Be alloy, which has a very similar thermal contraction as a typical Bi2212 composite. With the Bi2212 wire specimen soldered to the spring, from room temperature to 5 K the differential thermal contraction between the two materials produces just a 0.01% compression on the sample.

Fig. 11 shows the normalized $I_c$(4.2K) as a function of applied strain for a representative set [14] - [16] of the several accurate studies on Bi2212 axial strain behavior that can be found in literature on wires from different manufacturers. The plot includes a Showa wire from the 2000s, a Supercon wire, and an OST-Bruker wire with Nexans powder. Note that the magnetic field in each study is different. Under tensile strain, the $I_c$ of all wires decreases slowly, at about -2% to -3% per percent strain, until a precipitous drop. However, the drop occurs at different strain values for each wire. For instance, the irreversible strain limit (the strain at which $I_c$ is permanently reduced to 95% of its original value) for the Showa wire was the largest at 0.5%, even if the background field of 30 T was the largest. In compression only the OST-Bruker was tested. The $I_c$ drops linearly by about 26% per percent strain, and the drop was shown to be irreversible from the start. When the strain is inverted from compressive to tensile, independently of the strain value at which this occurs, the $I_c$ behavior resumes as if the sample were in tensile mode. Although all Bi-2212 is made with the PIT technique, this plot indicates that strain behavior depends on the specific technology used by each manufacturer.

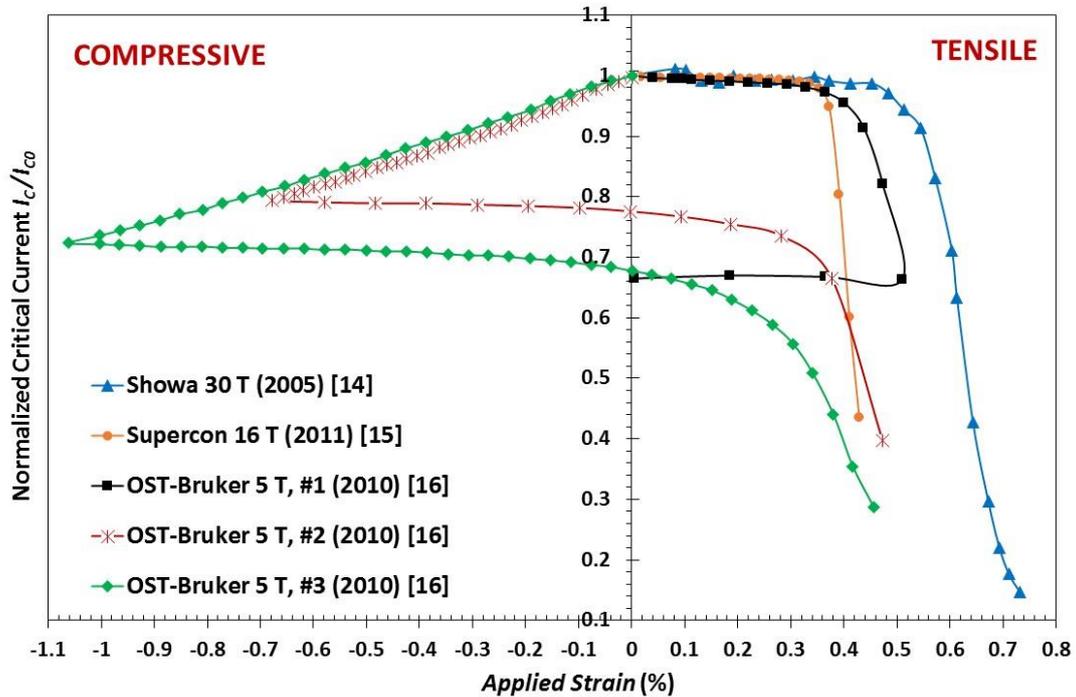

Figure 11: Normalized $I_c$(4.2 K) as function of applied longitudinal strain for a number of Bi2212 wires and magnetic fields [14] - [16].

## $I_c$ Sensitivity to Transverse Load

Transverse stress, as the largest stress component in accelerator magnets, can damage brittle Bi2212 coils. To determine $I_c$ sensitivity to transverse pressure, electro-mechanical tests are typically performed on either cables or encased wires. Transverse pressure studies are made by applying pressure to impregnated Rutherford cable or wire samples and testing their transport current at several magnetic fields. Strain sensitivity, as previously indicated, increases with magnetic field. There are two components of the critical current degradation: a reversible component, which is fully recovered when removing the load, and an irreversible component. The latter is permanent. The irreversible limit is defined as the pressure leading to a 95% recovery of the initial $I_c$ after unloading the sample. Institutions where transverse pressure measurements can be performed include FNAL, the University of Twente, the National High Magnetic Field Laboratory (NHMFL), CERN, and the University of Genève. The $I_c$ degradation strongly depends, as previously seen, on the Bi2212 wire technology, but also on sample preparation and setup design. The former has an impact on possible stress concentrations; the latter determines the sample's actual stress–strain state.

Data for transverse pressure testing of Bi2212 Rutherford cables are presented in Fig. 12 for a Bi2212 sample of older generation (i.e. Nexans powder) heat treated at 1 bar pressure, together with transverse pressure data for other commercial superconductors. The FNAL Transverse Pressure Insert (TPI) measurement system is a device to test critical current sensitivity of impregnated superconducting cables to uniaxial (plane stress) transverse pressures up to about 200 MPa. It should be noted that this device produces the effect of uni-axial and not multi-axial strain, since the experimental setup allows for the sample to expand laterally. This produces larger strain values on the cable sample than for instance on a laterally constrained one. Fig. 13, left, shows a $Nb_3Sn$ sample before impregnation. Fig. 13, center, shows a schematic of the current carrying wire in between adjacent Cu dummy wires in the cable package. Fig.

13, right, shows the sample after testing, as mounted to the fixture with the bottom pressure plate removed. For Bi2212 sample testing, Ag cable dummies are used. This transverse pressure test will have to be repeated with state-of-art Bi2212 conductor, and will have to be also performed for samples heat treated at 50 bar gas pressure.

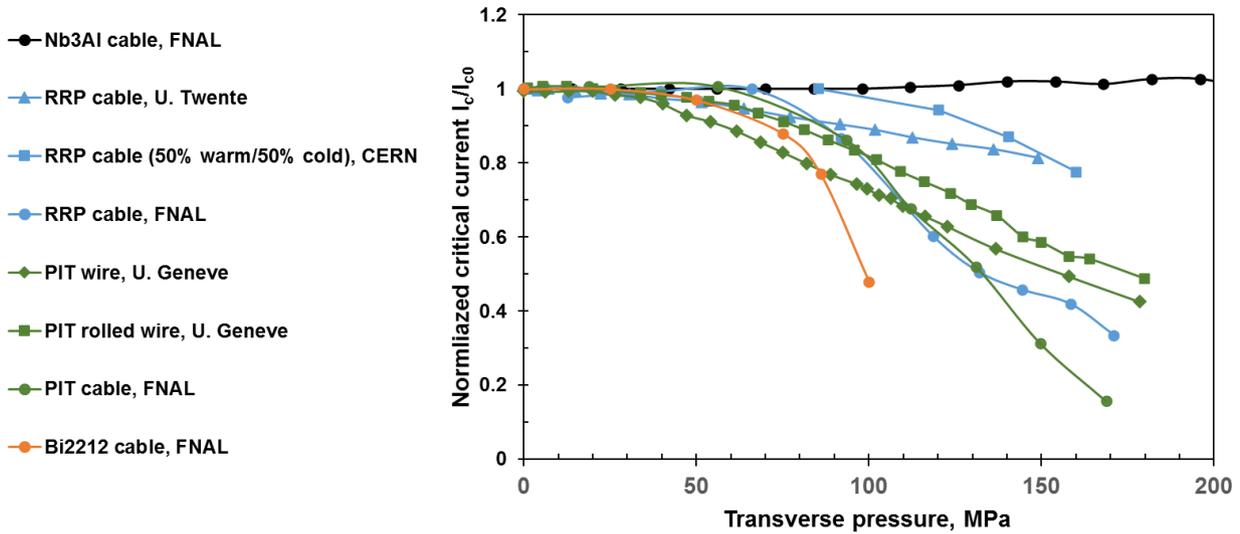

Figure 12: Normalized $I_c$ at 4.2 K vs. transverse pressure on Rutherford cable face or on encased wire for epoxy impregnated samples. All the $I_c$ data correspond to 12 T, except for the PIT round and rolled wires which were measured by U. Genève at 19 T.

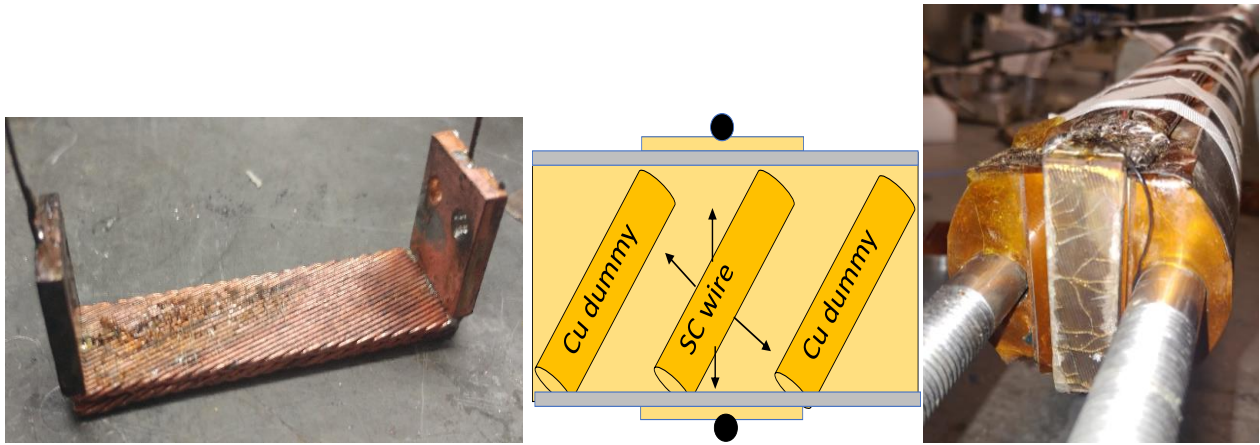

Figure 13: Left: Single $Nb_3Sn$ strand embedded in the bottom dummy Rutherford cable blank of a two-layer Cu cable blank stack. Soldered Cu tabs are shown; these are for increasing current carrying capacity near the bends and mechanical stabilization. The brown residue is from soldering flux and is removed before epoxy vacuum impregnation. Center: Schematic of current carrying wire within Cu dummies. The voltage taps (black circles) are located on the Cu tabs. Right: Sample after testing, as mounted to the fixture with the bottom pressure plate removed.

Because of the greater availability and accuracy of axial strain data, it would be profitable to establish a strain equivalent model that produce a total equivalent strain from transverse pressure configurations. An attempt was herein made in Appendix A using the formulae for the Von Mises equivalent strain, for both cases of laterally constrained and non-constrained cable sample.

In Fig. 14 the Von Mises strain associated to the pressure data points from the experiment in Fig. 12 was calculated using Eq. A4, with a Poisson ratio ν=0.3, and a Young modulus E = 20 GPa. The pressure applied in the experiment is shown on the primary y axis, and the associated Von Mises equivalent strain on the secondary y axis, both as a function of the measured normalized $I_c$ reduction. Because the cable was made with Bruker-OST wire, the compression data from NIST for two Bruker-OST samples [16] are also shown in the plot with inverted axis with respect to Fig. 12. From these very limited data, it appears that the Von Mises equivalent strain of the cable sample under compression reasonably represents the tensile strain behavior of the wire up to the $I_c$ drop. After the drop, it is instead the compressive strain behavior of the wire that seems to control the Von Mises strain. Investing in more transverse pressure experimental data is vital, both for magnet design and to improve equivalent strain models.

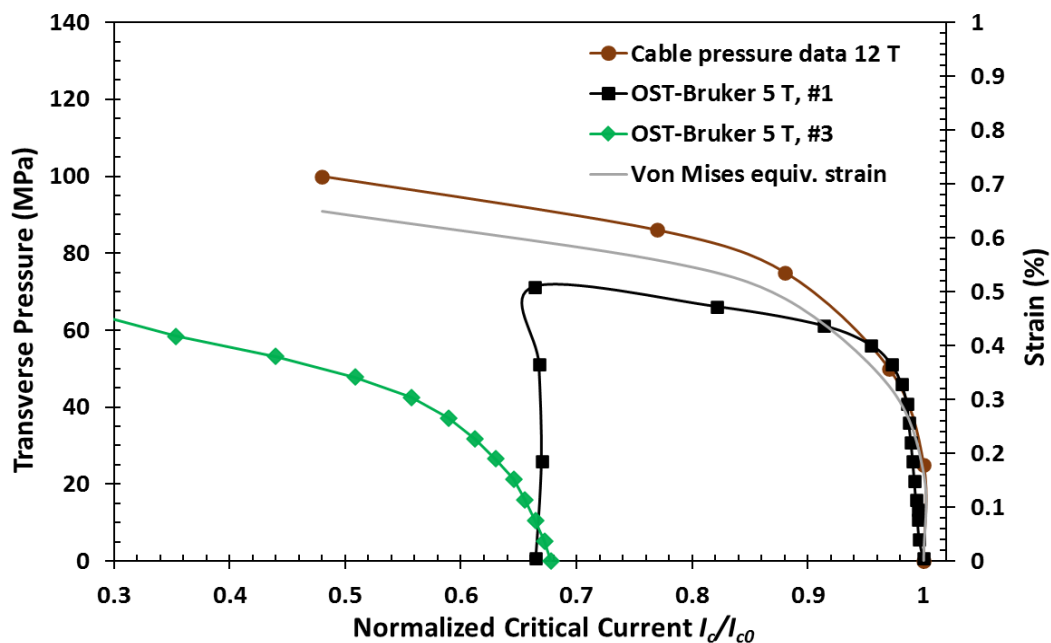

Figure 14: Applied transverse pressure applied is shown on the primary y axis as a function of the measured normalized $I_c$ reduction for the Bi2212 sample in Fig. 12. On the secondary y axis representing strain are the OST-Bruker tensile strain data [11] for samples #1 and #3, and the calculated Von Mises equivalent strain.

### 5.3. Cable Characterization

The following wire procurement and qualification plan, to evaluate and select the most adequate conductors, is the standards for the ultimate test of the conductor in magnet models:

- Procurement of billets from industry in sufficiently large quantities to verify production quality.
- Qualification of round and deformed strands for $I_c$, n-value, flux jump stability, $d_{eff}$, RRR, and $I_c$ sensitivity to transverse pressure.
- Rutherford cable development and fabrication.
- Characterization of strands extracted from cables and/or cable test.

**Characterization of Deformed Wires and Strands Extracted from Cables**

Superconducting wires are flat-rolled to increasing deformations to systematically study such dependence of their properties. Wire deformation is defined as (d-t)/d, where d is the strand diameter and t is the thickness of the deformed strand. After heat treatment, size variations of round and deformed samples are measured, and their transport properties characterized at 4.2 K and at magnetic fields up to 15 T. Schematics of the test setup are shown in Fig. 15. For this measurement the $I_c$ is determined on the innermost voltage taps using a 1 µV/cm criterion. An example of $I_c$ and n-value results from this type of study is shown in Fig. 16 for three Bi2212 wires made by Bruker-OST, with wires A and B having Nexans powder as precursor, and wire C having Nexans granulate [31]. The n-value is a measure of wire inhomogeneity. The conclusion for this study was that at the highest tested deformation of 60%, the $I_c$ of the three wires was either close to its value in the undeformed sample, or even larger for one of the wires made with powder. The n-value of all three wires had also recovered their original undeformed strand values, as if filament separation self-corrected as the wire was further rolled to smaller thicknesses. Also, wire C made with granulate precursor was the most homogenous wire at all deformation levels. The wires made with powder showed an even stronger non-monotonic behavior in $I_c$ and in the n-value.

The results from this study, together with accurate measurements of variations in wire diameter before and after heat treatment, which showed expansion [31] consistently with gas developing during the reaction and getting trapped in the pores of the filament bundles, led the VHFSM collaboration to using the overpressure processing heat treatment OPHT.

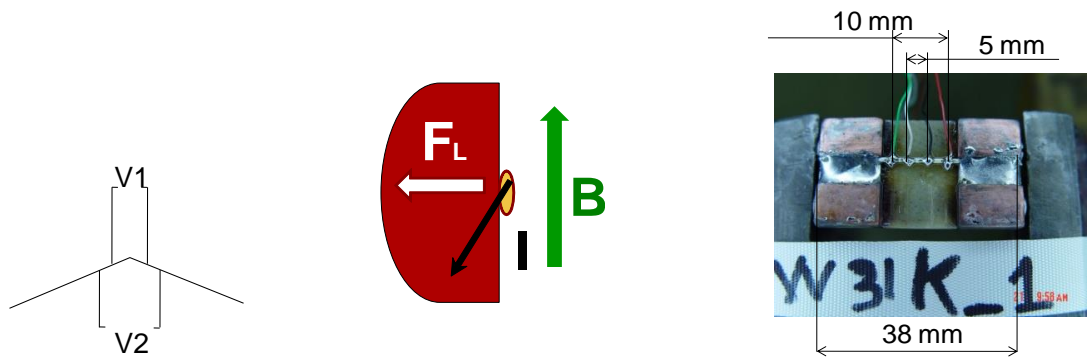

Figure 15: Left: The $I_c$ is determined on the innermost voltage taps. For extracted strands, the voltage taps include a cable edge. Center: Rolled wire sample with magnetic field parallel to its width. Right: $I_c$ test setup.

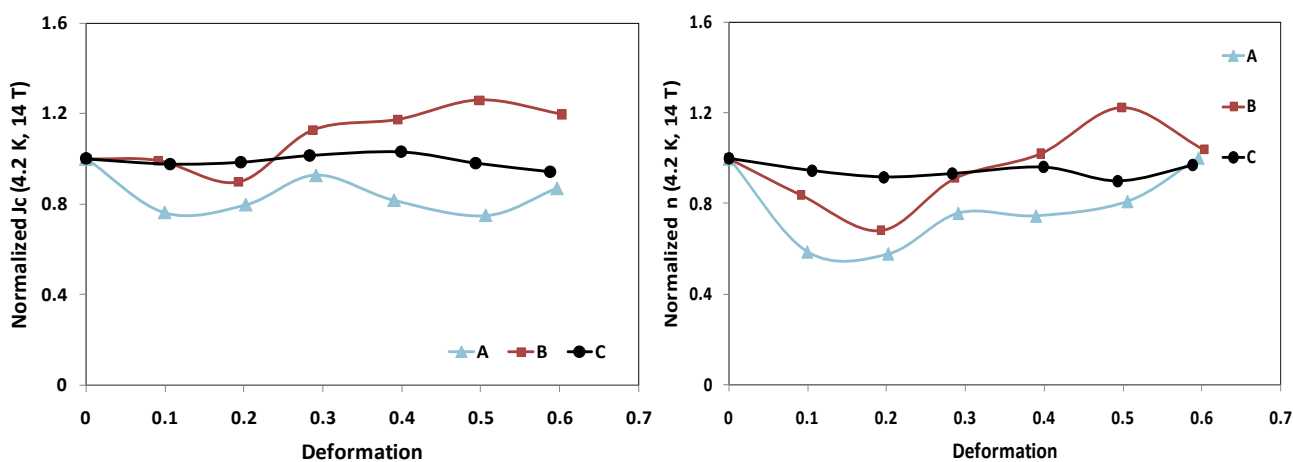

Figure 16: Left: $J_c$(4.2 K, 14 T) of a deformed strand normalized to the $J_c$(4.2 K, 14 T) of the virgin wire as a function of strand deformation for three different Bi2212 wires A, B and C by Bruker-OST. Right: n-value(4.2 K, 14 T) of a deformed strand normalized to the n-value(4.2 K, 14 T) of the virgin wire as a function of strand deformation for three different Bi2212 wires by Bruker-OST [31].

Strands are extracted from cables before heat treatment and tested after reaction as a means to characterize a cable. During cable development, the $I_c$ and n-value at a given magnetic field, and other properties of the extracted strand can be plotted against cable compaction and/or cable mid-thickness at each fabrication step. $I_c$ degradation values < 10% are typically desired. Assuming a uniform current distribution among the strands, cable $I_c$ is defined as the strand $I_c$ multiplied by the total number of strands in the cable. For $Nb_3Sn$, the most sensitive indicator of cable damage are increased flux jump instabilities in extracted strands, for which a macroscopic measurement is the instability current $I_S$ at low magnetic fields. Presumably Bi2212 wires will have a different main indicator. Another important use of extracted strand tests is to include samples with the magnet coil during high temperature reaction as witnesses of the process. Some representative results from [26], [32] were plotted together in the graph of Fig. 17. A theoretical cable packing factor of 78.5% is used for the virgin wire, i.e. cable made of undeformed round strands. All the results shown in the plot are for Bruker-OST wires with Nexans powder.

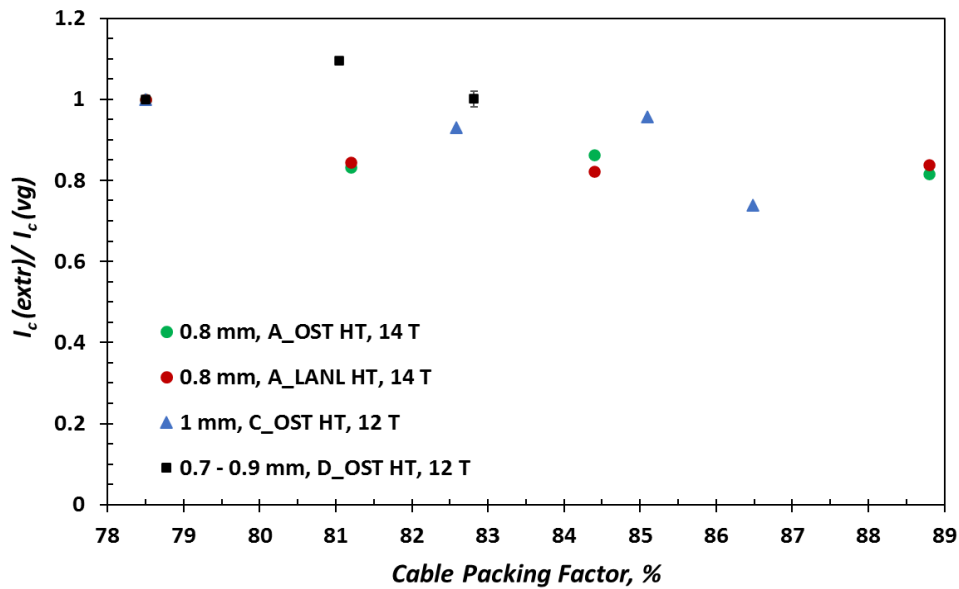

Figure 17: Normalized $I_c$ (4.2 K) as a function of cable packing factor for extracted Bi-2212 strands of type A, C and D, heat treated at OST and/or at LANL [26], [32].

The method of flat-rolling wires can also be used to reproduce the deformation seen in a Rutherford cable. Strand deformation values can be measured from Rutherford cable cross-sections by using the formula $(d_{max} - d_{min})/d$, where $d_{max}$ is the maximum diameter of a strand and $d_{min}$ its smallest. Fig. 18 shows measurements of strand deformation as a function of strand location in a Bi2212 cable 1.49 mm thick and 10.38 mm wide, made of 25 strands of 0.8 mm in diameter, with 84.4% packing factor. The cable edges are located at strand location No. 1, 12-13, and 25 [32].

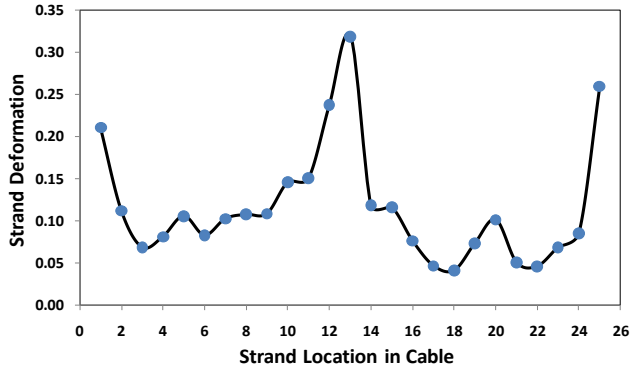

Figure 18: Strand deformation in Bi2212 cable 1.49 mm thick and 10.38 mm wide, made of 25 strands of 0.8 mm in diameter (packing factor = 84.4%), as function of strand location. Maximum deformation is 0.32 on one edge, average is 0.12 ± 0.07 [32].

**Cable Test at Self-field**

Bi-2212 Rutherford cables can be tested at self-field with a superconducting transformer equipped with a Rogowski coil to measure the secondary current transported by the cable sample [33]. Fig. 19, left, shows a schematic of the voltage taps location on the cable sample. Fig. 19, right, shows an example of Bi2212 cable test [32], for which the cable critical surface obtained from extracted strand test results is compared with the transformer load line to produce the expected cable performance, or short sample limit predictions. This Bi-2212 Rutherford cable reached 90% of the current predicted by measuring extracted strands. The cable geometry is the same as in Fig. 18, and the cable was heat treated at LANL, together with its witness samples, which were strands extracted from this same cable. The cable was non-insulated and non-impregnated and the average $I_c$ obtained over three $I_c$ tests using a 10 μV/cm criterion was 7315 A ± 128 A. Its n-value was 11.5 ± 0.2. To calculate the expected current, data from the witness samples were used. Fig. 19, right, shows the $I_c$ as a function of field for the best and worst performing samples, corrected for their self-field. The same 10 μV/cm criterion was used as for the cable. The expected cable $I_c$ was calculated by intersecting the lowest curve with the load line of the transformer secondary, which represents the cable self-field as a function of current. The intersection was found at 8106 A.

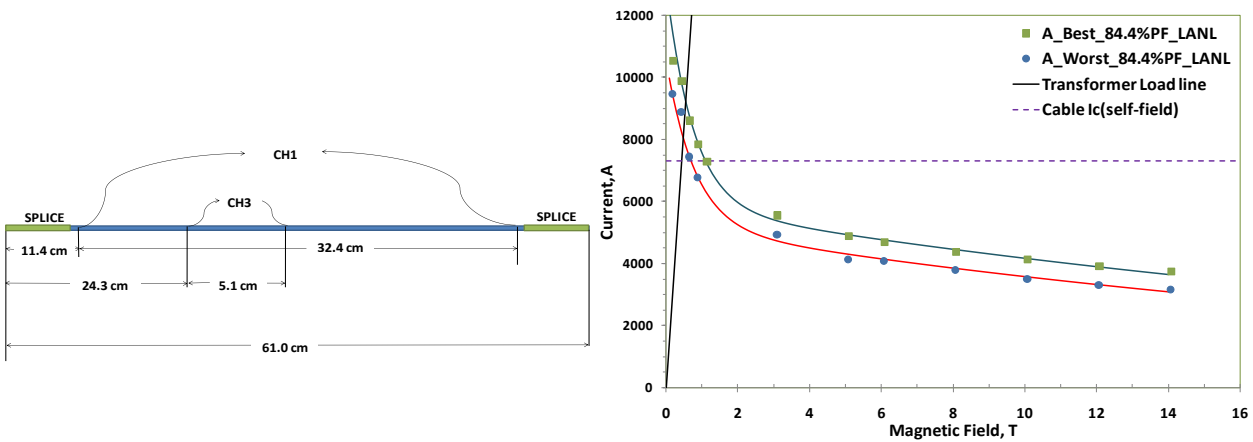

Figure 19: Schematic of cable test and voltage tap location. Right: Short sample limit predictions of Bi-2212 Rutherford cable Ic based on extracted strand tests and the transformer load line [32].

**Cable Test at Field up to 15 T**

$I_c$ measurements of Rutherford cables are performed using a test fixture with bifilar sample and superconducting transformer that operates in the 14T/16T solenoid of a Teslatron system by Oxford Instruments [34]. For $Nb_3Sn$ cable sample, a stainless steel sample holder is used both for reaction and test. Inconel holders have to be fabricated for reaction and testing of Bi2212 cables.

## 6. CONCLUSIONS

The Bi2212 dipole insert program started at Fermilab a couple of years ago is to test and develop the technology of HTS inserts based on Bi2212 Rutherford cable and cos-theta coil configuration. On paper, the potential reach for the maximum magnetic field in existing or planned $Nb_3Sn$ outserts is close to 20 T, thanks to the progress realized in wires' critical current density. To achieve the Bi2212 potential in accelerator magnets, however, a number of technological challenges still have to be faced. These for instance include the need to design billets that are adequate for Rutherford cabling; developing insulation processes and materials that prevent leaks, which reduce transport current and increase the risk of shorts; control and limit Bi2212 coils' stresses and strains; reconsider the Split Melt Process (SMP) to lower costs and simplify the processing.

## APPENDIX A. Von Mises Equivalent Strain

Transport properties measurements for Rutherford cables under load are based on the principle of applying a transverse pressure on a cable sample and then measuring its critical current $I_c$. We can think of at least two different boundary conditions for the cable, which may produce different current density performances: plane stress with free sides, and plane stress with constrained sides (Fig. A1).

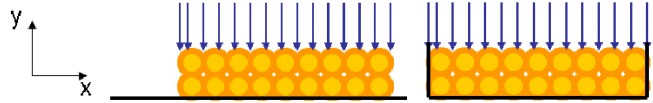

Figure A1: Uni-axial case A, free sides (left), and multi-axial case B (right).

The first load case, uni-axial case A, represented in Fig. A1 (left), has $\sigma_{yy} = -p$ and $\sigma_{xx} = \sigma_{zz} = 0$. It can therefore be represented by the matrix:

$$\begin{pmatrix} 0 & 0 & 0 \\ 0 & -p & 0 \\ 0 & 0 & 0 \end{pmatrix}.$$

The strain tensor for case A is therefore given by:

$$\begin{pmatrix} \nu\frac{p}{E} & 0 & 0 \\ 0 & -\frac{p}{E} & 0 \\ 0 & 0 & \nu\frac{p}{E} \end{pmatrix} \quad (A1),$$

where $\nu$ is the Poisson coefficient and $E$ the Young modulus.

The second load case, multi-axial case B, represented in Fig. A1 (right), has $\sigma_{yy} = -p$, $\sigma_{xx} = -\nu p$ and $\sigma_{zz} = 0$. To generalize this load case, we suppose to give a preload on the $x$ direction equal to $\varepsilon_{xx0}$. The strain tensor for case B can then be represented by:

$$\begin{pmatrix} \varepsilon_{xx0} & 0 & 0 \\ 0 & (\nu^2-1)\frac{p}{E} - \nu\varepsilon_{xx0} & 0 \\ 0 & 0 & \nu(1+\nu)\frac{p}{E} - \nu\varepsilon_{xx0} \end{pmatrix} \quad (A2).$$

The Von Mises equivalent strain $e_{eq}$, defined by the following formulae, is a good candidate for a strain that might be representative of a cable transverse pressure setup:

$$\varepsilon_{eq}^2 = \frac{1}{2}\left[(\sigma_{xx}-\sigma_{yy})^2 + (\sigma_{yy}-\sigma_{zz})^2 + (\sigma_{zz}-\sigma_{xx})^2 + 6(\sigma_{yz}^2 + \sigma_{zx}^2 + \sigma_{xy}^2)\right];$$

$$\varepsilon_{xx} = \frac{1}{E}\left(\sigma_{xx} - \upsilon(\sigma_{yy}+\sigma_{zz})\right);$$

$$\varepsilon_{yy} = \frac{1}{E}\left(\sigma_{yy} - \upsilon(\sigma_{xx}+\sigma_{zz})\right);$$

$$\varepsilon_{zz} = \frac{1}{E}\left(\sigma_{zz} - \upsilon(\sigma_{xx}+\sigma_{yy})\right);$$

$$\varepsilon_{xy} = \frac{1}{2G}\sigma_{xy}; \quad \varepsilon_{xz} = \frac{1}{2G}\sigma_{xz}; \quad \varepsilon_{yz} = \frac{1}{2G}\sigma_{yz}.$$

By applying the fomulae above, using for case A the strain values from Eq. A1, and for case B the strain values from Eq. A2, and assuming $\varepsilon_{xx0} = 0$, the following equivalent Von Mises strains are obtained:

$$\varepsilon_{eqA} = \frac{p}{E}(\nu+1), \text{ and} \quad (A3)$$

$$\varepsilon_{eqB} = \frac{p}{E}\sqrt{(\nu^3+1)(\nu+1)} \quad (A4).$$

## Acknowledgments

The author would like to thank Alexander V. Zlobin, Igor Novitski, Dan Turrioni and Alessio D'Agliano, Fermi National Accelerator Laboratory, Batavia, IL 60510, who provided some of the figures for this paper, and Jianyi Jiang, Applied Superconductivity Center, Florida State University, Tallahassee, FL 32310, for kindly performing heat treatment of Bi2212 samples at 50 bar gas pressure, and the many physicists, engineers, technicians and graduate students that contributed to these results.